\documentclass[alpha-refs]{wiley-article}

\usepackage{siunitx}
\usepackage{tikz}
\usepackage{pgfplots}

\usepackage{caption}
\usepackage{subcaption}
\usepackage{calc}
\usepackage{graphicx}
\usepackage{amsmath}

\newlength{\eqwidth}
\newlength{\arrwidth}
\usepackage{lineno}

\papertype{Research Article}

\title{A hypothesis on ergodicity and the signal-to-noise paradox}

\abbrevs{NAO, North Atlantic Oscillation; RPC, ratio of predictable components; PDF, probability density function; GloSea5, Met Office Global Seasonal Forecast System, version 5.}

\author[1\authfn{1}]{Daniel J. Brener}

\affil[1]{The Higgs Centre for Theoretical Physics, The University of Edinburgh, Edinburgh, EH9 3FD, UK}

\corraddress{Daniel J. Brener, The Higgs Centre for Theoretical Physics, The School of Physics \& Astronomy, The University of Edinburgh, James Clerk Maxwell Building Room 4301, Kings Buildings, Peter Guthrie Tait Road, Edinburgh, EH9 3FD, UK}
\corremail{daniel.brener@ed.ac.uk}

\fundinginfo{Science and Technology Facilities Council, UK.}

\runningauthor{D. J. Brener}

\begin{document}

\maketitle
\begin{abstract}
This letter raises the possibility that ergodicity concerns might have some bearing on the signal-to-noise paradox. This is explored by applying the ergodic theorem to the theory behind ensemble weather forecasting and the ensemble mean. Using the ensemble mean as our best forecast of observations amounts to interpreting it as the most likely phase-space trajectory, which relies on the ergodic theorem. This can fail for ensemble forecasting systems if members are not perfectly exchangeable with each other, the averaging window is too short \textcolor{black}{and/or} there are too few members. We argue these failures can occur in cases such as the winter North Atlantic Oscillation (NAO) forecasts due to \textcolor{black}{intransitivity} or regime behaviour for regions such as the North Atlantic and Arctic. This behaviour, where different ensemble members may become stuck in different relatively persistent flow states \textcolor{black}{(intransitivity)} or multi-modality \textcolor{black}{(regime behaviour)}, can in certain situations break the ergodic theorem. The problem of non-ergodic systems and models in the case of weather forecasting is discussed, as are potential mitigation methods and metrics for ergodicity in ensemble systems.

\end{abstract}
\newpage
\section{Introduction}
The equations governing the atmosphere were shown in the seminal papers of \cite{lorenz1963deterministic} and \cite{lorenz1969predictability} to exhibit sensitive dependence on initial conditions, commonly known as chaos. Given the impossibility of perfectly knowing these conditions, predictability of the atmosphere is inherently limited, even with a perfect model. To address this uncertainty, the method of ensemble weather prediction was developed, see \cite{epstein1969stochastic}. Employing multiple realisations of a numerical model with tiny variations in initial conditions, these small perturbations give rise to divergent predictions due to chaos. The average value of an observable over several ensemble members, can sometimes be interpreted as the best estimate of the said state of the observable at or over a particular time. The mean and standard deviation of predictions across ensemble members provide estimates of an observable's state and uncertainty, despite the deterministic equations underlying the models, as discussed in \cite{murphy1986experimental} and \cite{palmer2006predictability}.

\textcolor{black}{\cite{Scaife2018} reviewed what has come to be known as a signal-to-noise paradox:} that atmosphere-ocean coupled climate and long range prediction ensemble models are better at predicting reality, than they are at predicting themselves. \textcolor{black}{This was first raised as possibility by \cite{kumar2009finite} and consequently demonstrated by \cite{Eade2014}.} They derived a statistic for comparing ensemble models to observations known as the ``ratio of predictable components" (RPC) for several atmospheric parameters,
\begin{equation}
    RPC^2 = \frac{r_{\bar{E}O}^2}{r_{\bar{E}E_i}^2},
\end{equation}
where $r_{\bar{E}O}$ is the Pearson correlation between the model ensemble mean and the observations, and $r_{\bar{E}E_i}$ is the average correlation between the model ensemble mean and a single ensemble member. It was found that the correlation between the ensemble mean and observations is often much greater than the average correlation between the ensemble mean and a single ensemble member ($RPC > 1$), referred to as an anomalous RPC. Hence, a signal-to-noise paradox - the model predicts reality better than it predicts itself. This has since been reported by many different groups such as \cite{Scaife2014} \cite{Stockdale2015}, \cite{Charlton-Perez2019}, \cite{Weisheimer2019} and \cite{Dunstone2023}.

Over the past few years a great deal of effort from the community has been put into finding different resolutions to the paradox, and different scenarios where it arises. \cite{https://doi.org/10.1029/2023GL103710} showed that for the North Atlantic Oscillation (NAO), the paradox may be interpreted as a probabilistically under-confident forecast for occurrences of high magnitude NAO. A possible explanation for this is that the model has reduced persistence of particular atmospheric regimes, especially in the Northern Hemisphere according to \cite{Strommen2018} and \cite{strommen2020jet}. In terms of model dynamics, weak atmosphere eddy feedback was suggested by \cite{scaife2019does} and \cite{hardiman2022missing}, or weak ocean-atmosphere coupling in models by \cite{osso2020development}. Early approaches to the paradox hoped that enhancing model physics could directly improve the correlation between the model and the ensemble mean, thereby increasing forecast skill. These improvements have not yet been achieved. A detailed summary of the current state of understanding and avenues to tackle the problem can be found in \cite{antje_report}. Taking a different angle, \cite{https://doi.org/10.1002/qj.4440} argued that in some cases, \textcolor{black}{that the} signal-to-noise paradox should not be considered paradoxical due to the assumption that the forecast error is related to the correlation of the ensemble mean with the observations, which is not necessarily always the case. It is in this more statistical direction that this Letter takes us.

We suggest that part of the paradox could be \textcolor{black}{due to a violation} of the ergodic theorem which is relied upon for the statistical moments comprising the RPC. This is because using the ensemble mean as our best forecast of observations amounts to interpreting it as the most likely phase-space trajectory, which relies on the ergodic theorem. We argue that \textcolor{black}{this can fail} in cases such as the winter NAO forecasts due to \textcolor{black}{intransitivity/multi-modality} which can in certain situations break the ergodic theorem. The ideas presented are not intended to resolve the paradox entirely as there are other contributing issues related to both modelling and the definitions used in the computation of the RPC (e.g. \cite{https://doi.org/10.1002/qj.4440}, \cite{Zhang2019}, \cite{https://doi.org/10.1029/2022GL100471} and \cite{https://doi.org/10.1002/qj.3413}). We hope this paper will act as a catalyst for others to consider how the ergodicity assumption might be tested more rigorously, and serve as a reminder of it.
\section{Ergodic theory}
\subsection{The ergodic theorem}
Consider a dynamical system, $x_t$ evolving with time $t$ on some phase space $X=\mathbb{R}^d$, where $d$ is the system dimension. Suppose there exists some observable $f(x_t) \in X$, which can be integrated such that a measure $\mu$ on $X$ is preserved. Let $T$ be a unique time-evolution operator such that an initial state $x_0$ can evolve to a discrete later time $x_l$ via $T^l x_0 = x_l,$ $l\in\mathbb{N}$. If $T$ preserves the invariant measure $\mu$, then $T$ is said to be ergodic. Our observable $f$ has a \textit{time mean of a single path} on $X$ going from an initial position $(x_0,t_0)$ to a later position $(x_l, t_l)$ given by,
\begin{equation}
    \bar{f}(x_0) =\lim _{l \rightarrow \infty} \frac{1}{l} \sum_{k=0}^{l-1} f\left(T^k x_0\right).
\end{equation}
For the same observable $f$, one can also find the \textit{space mean},
\begin{equation}
    \bar{f}(x_{t^\prime}) = \int \mu f(x_{t^\prime}) dx
\end{equation}
at an instant in time $t^\prime$. Under certain conditions on the system, the ergodic theorem holds that: the time and space means are the same for all possible initial conditions $x_0$. This is also known as the ergodic hypothesis and was proved, along with the existence of the averages individually, by \cite{1bd409ba-8905-3534-9c6f-e5a5e359481e}. See \textcolor{black}{\cite{MoulinOllagnier1985}} for a formal approach to the subject. In the next section we will see how this works for ensemble models.

\subsection{Application to ensemble modelling}
\textcolor{black}{In numerical weather prediction, $x_t$ is the state of the atmosphere at time $t$, evolving in some phase space $R^d$, where $d$ is very large. The operator $T$ represents the evolution of the atmosphere forward in time according to all the equations of physics. The phase space of the various atmospheric/climatic states is some strange attractor, $\mathcal{A}$ see \cite{ChaosStrangeAttractorsandWeather} for an introduction. $T$ takes us between states on $\mathcal{A}$, whilst preserving $\mathcal{A}$. If $T$ is chosen to be ergodic, then $T$ will visit all parts of $\mathcal{A}$ over time; for a rigorous treatment see \cite{RevModPhys.57.617} part 2E.}

The ensemble mean of an observable, $E(f(x_t))$ at a time $t$ is,
\begin{equation}
\begin{split}
    E(f(x_t)) = \frac{1}{n} \sum_{i=1}^{n} f^i(x_t)
\end{split}
\label{eqn:ensemble_mean}
\end{equation}
where $f^i(x_t)$ is the observable from the $i^{th}$ member of the ensemble of size $n$. The significance of the ensemble size, $n$ has been well studied on all temporal and spatial scales; see \cite{palmer2006predictability} and \cite{https://doi.org/10.1002/qj.3387}. Under a ``signal + noise" model, the idea is that forecast skill grows with $n$ due to the suppression of unpredictable noise.

In using the RPC metric, we want to interpret the ensemble mean as being the \textit{most likely trajectory} in the phase space $X$, and hence our best guess for what the observations will do. This is an ergodicity assumption as we want to equate the space average (ensemble mean) with the most likely trajectory (time average). An ensemble is initialised as a set of perturbed initial states $\{x_0^i\}$. There must exist a unique time-evolution operator $T^k_i$ for each member that evolves each initial state over a time $\Delta T \equiv T_{k=l} - T_{k=0}$. So our ensemble mean can be given by,
\begin{equation}
\begin{split}
    E(f(x_l)) & = \frac{1}{n \cdot l} \sum_{i=1}^n \sum_{k=0}^{l-1} f(T^k_i x_0^i)
\end{split}
\end{equation}
In other words, running our ensemble simulation is equivalent to computing the evolution of the initial states by the operator $T$. Since $T$ is \textcolor{black}{unique}, it is more convenient to think of each ensemble member as having their own continuous time-dependant probability density function (PDF), $P^i(x_t)$, that evolves according to unique $T_i^l$, so $T^l_iP^i(x_0) = P^i(x_l)$. This gives the ensemble mean as
\begin{equation}
\begin{split}
    E(f(x_t)) & = \frac{1}{n \cdot \Delta T}
    \int_{T_0}^T{ f^1(x_t)P^1(x_t) + f^2(x_t)P^2(x_t) + ... + f^n(x_t)P^n(x_t)dt},
\end{split}
\label{ens_mean}
\end{equation}
so that the observable as measured in each member is weighted by the evolving PDF unique to each member. This average will only represent the most likely trajectory if the following conditions hold:
\begin{alignat*}{3}
    \lim n&\rightarrow &&\infty; \text{ infinite ensemble size},\\
    \lim \Delta T &\rightarrow &&\infty; \text{ infinitely long averaging window},\\
    \left. \frac{d P^i}{dt}\right|_{t=\Delta T, \forall i} &\scalebox{1.4}[1]{$ =$} &&0; \text{ distribution function for each ensemble member is stationary as } T^i \text{ must be unique over period } \Delta T.
\end{alignat*}
The final condition is equivalent to saying that the ensemble members must be \textit{exchangeable}, which is a common assumption for ensemble prediction systems and will be discussed in a later section. We will next interrogate these concepts using a Galton Board for Gedankenexperimente.

\subsection{Ergodicity in the Galton Board: winding a classic model}

\begin{figure}[htb!]
  \centering
  \begin{subfigure}[b]{0.43\linewidth}
    \includegraphics[width=\linewidth]{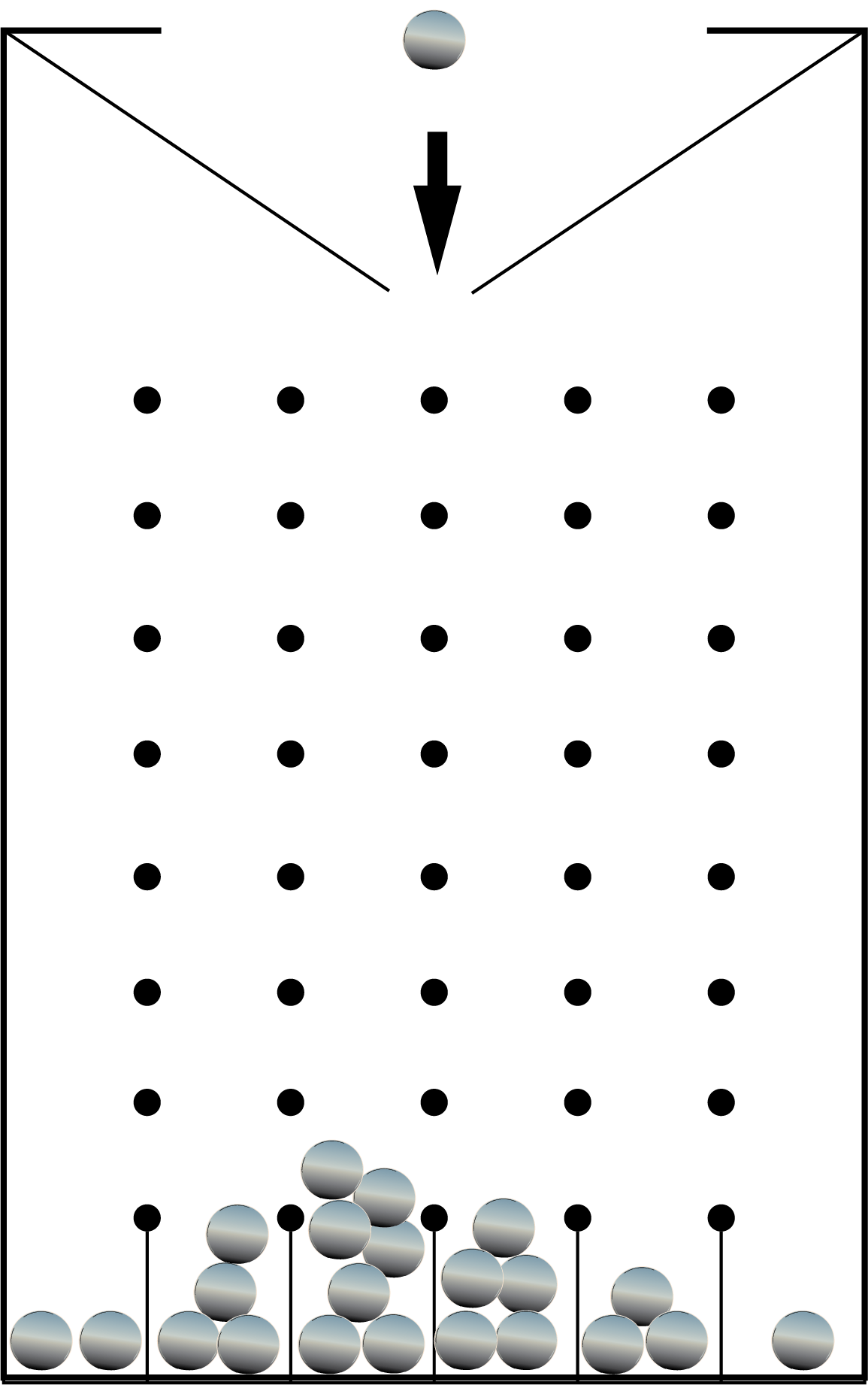}
    \caption{Fixed length classic Galton board.}
    \label{fig:galton1}
  \end{subfigure}
  \hspace{1pt}
  \begin{subfigure}[b]{0.55\linewidth}
    \includegraphics[width=\linewidth]{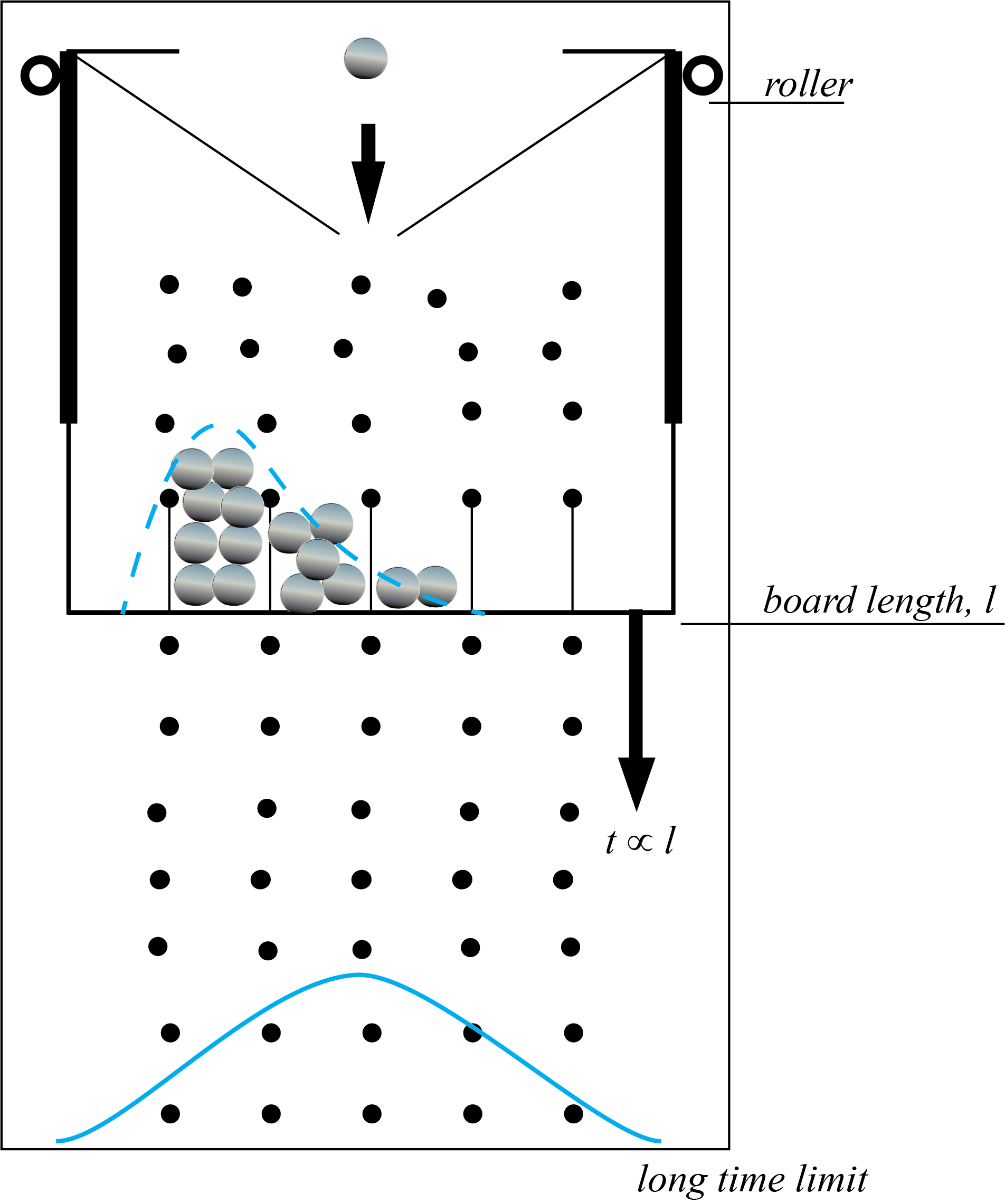}
    \caption{Galton board with changeable length and exaggerated peg-misalignment.}
      \label{fig:galton2}
  \end{subfigure}
  \caption{Illustration of our two Galton boards. Not drawn to scale. The right-hand board has adjustable length using a roller. Adjusting the length corresponds to adjusting the time the balls have to explore the system. The longer the board, the longer the time the balls spend in the system and the better representative the ensemble mean and variance will be of the underlying PDF, even if the space is multi-modal. Blue solid line in Figure (b) indicates the Gaussian outcome of the long-board (time) limit, in contrast to distribution of balls (blue dashed) with some fixed short board length (time). Note that the distribution of balls is not instantaneous, but an average over a number of different balls (ensemble members) over a time proportional to the length they are allowed to drop.}
\end{figure}

A Galton board is a device invented by \cite{galton1889natural} that consists of a vertical board with an array of pegs. At the top, there is an entry point from which small balls, typically marbles, are released. The Galton board is often used to visually illustrate how randomness, combined with a large number of trials (balls), leads to a Gaussian distribution. In meteorology, it has been used to teach the principles of ensemble forecasting and has also been used to develop conceptual arguments in this field. We illustrate a normal Galton board in Figure \ref{fig:galton1} where after multiple individual balls we begin to approximate a Gaussian.

For our purposes, each individual ball trajectory represents the time evolution of a single ensemble member and the statistical distribution of the final positions represents the ensemble average for the model. As one observes more and more balls passing through the board, the distribution of final positions at the bottom converges to a pattern that resembles a Gaussian distribution. This convergence is analogous to the ergodic theorem, where the time average (individual ball's trajectory) converges to the ensemble average (statistical distribution of final positions) over a large number of trials. If we imagine an \textit{infinitely long board}, the long-time average trajectory for a single ball will be straight down; equivalent to the position of the ensemble mean. Its also worth noting from \cite{2000AIPC..511..144K} and \cite{Hoover1992-az} that the classic Galton board has been found to possess an ergodic strange attractor, and from \cite{judd2007galton} that it is not simply a random-walk phenomena. Hence it has useful parallels to weather and climate prediction.

We can play with this idealised ensemble model to examine what can happen if the ergodic assumption fails. The way this can happen is if the system contains multi-modality, in other words, there exist potential wells via certain trajectories into which the balls can stumble from which they are less likely to recover in a certain time-frame. This does not have to be dependant on the different initial conditions. In simulations of Galton boards such potential wells have been found to exist, see \cite{ahmed2022dynamics}.

Consider a Galton board of \textit{finite length} where we are able to modify this length using a roller, thus change the length of time that the balls must spend in the system. Now let us suppose that the gaps between the pegs are not identically uniform, so at certain points the balls path is impeded slightly more. Locally these act as potential wells, creating multi-modal features. The longer we make the board, the more time the balls will have to explore, enter and escape these potentials. \textcolor{black}{With a sufficiently long board (time), the resulting distribution, on average, will be our Gaussian again.} However, the shorter the board, the more the multi-modal features dominate the picture. One can imagine all kinds of situations that could occur, such as where the balls get stuck in roughly the same region of the board and so do not explore all the possible trajectories adequately. The longer we make the board, hence the longer the time, the more Gaussian our distribution of balls will become. We also need to use enough balls (large enough ensemble) to explore the board adequately. The key point is that the underlying distributions for the space/ensemble and time averages will be different when ergodicity fails.

Our model can also be extended to incorporate external forcings. For instance a light fan blowing across the board would shift the distributions, similar to the conceptual model of \cite{ANonlinearDynamicalPerspectiveonClimatePrediction}. The main difference between our Galton board and the picture of \cite{ANonlinearDynamicalPerspectiveonClimatePrediction} is that the latter constrains the system to a bimodal structure which is examined in the long time-limit, whereas our system allows us to imagine any number of modes (depending on the board ``resolution" and complexity) and our roller lets us examine long and short time limits. One can also think about exchangeability, as if one were to drop different kinds of balls (e.g. size/bass/stickiness), the final distributions would not make sense as a predictor for any one particular ball.

We can now see that without enough balls and or if the averaging time is too short, the ergodic theorem is not satisfied. In the case of the NAO, which evolves roughly daily, \textcolor{black}{then for a 90-day season in which the flow exhibits multimodality, 90 days will not always be sufficient to explore the complete phase space.} This is because such ensemble statistics reflect only those states the system has visited, highlighting the impact of daily variations on the dynamics that govern long-term averages.

\section{Failure of ergodicity and interpretation of the RPC}
We will now examine how failure of the ergodic assumption could lead to anomalous RPC values and hence the signal-to-noise paradox. To do this we draw on the statistical framework developed by \cite{https://doi.org/10.1002/qj.4440}. Let $V(e)$ and $V(o)$ be the respective variances of the ensemble and the observations, then it has been shown that,
\begin{equation}
\sqrt{\frac{V(e)}{V(o)}}<r_{\bar{E}O} \implies RPC > 1,
\end{equation}
where $r_{\bar{E}O}$ is the Pearson correlation between the observations and the ensemble mean. \cite{https://doi.org/10.1002/qj.4440} showed that when we have situations of small correlations, something highly likely for seasonal forecasting, then even minor differences between the observations and ensemble variance gives rise to a signal-to-noise paradox via the above inequality.

We make the conjecture that failure of ergodicity in an ensemble can in some cases lead to these differences. If the ergodic assumption fails for the mean, as defined by equation \ref{ens_mean}, then it also fails for \textit{all} statistical moments, including the variance $V(e)$. Mathematically, this is a strong statement as it renders such a variance as meaningless, as without ergodicity, the distribution will be different from sampling either from a long time-series of a single member or a complete ensemble. Like in our Galton board Figure \ref{fig:galton2}, the long-time distribution is a Gaussian, but with too short a board we are more likely to get some skewed distribution. The magnitude of the difference is something which can only be determined by direct computation of the distributions. However, \textcolor{black}{simply failing to respect ergodicity} does not mean that $V(e)$ should systematically underestimate $V(o)$.

Let us try a thought experiment with our windable Galton board in figure \ref{fig:galton2}. Suppose the observational reality is that the ball falls straight down, round a zero-mean Gaussian. We could conceive of a situation where our ensemble of balls gets stuck in a potential well such that they clump together relatively close to the centre. This would result in $V(e)<V(o)$ with a large $r_{\bar{E}O}$, hence an anomalous RPC value. From the weather regime perspective of the NAO, if one has multiple persistent regimes then a single 90 day season \textcolor{black}{will not always be sufficient} for a finite ensemble to adequately sample the space, necessary for the ergodic theorem. The members will appear to become ``stuck" in certain regimes. Consequently, each member on average then underestimates the total variance of a given season, leading to $V(e)<V(o)$.

The spatial inhomogeneity of the anomalous RPC values (see e.g. \cite{Weisheimer2019} for detailed study), especially over the North Atlantic and Arctic as reported by \cite{https://doi.org/10.1002/asl.1212} strongly suggest multimodality is involved, as these regions correlate with well-studied multi-modal behaviour. \cite{Strommen2018} point out that a multi-modal or regime approach provides an alternative perspective on the paradox. Using a toy bimodal Markov-chain model, they obtain anomalous RPC values when the persistence of the modes is underrepresented. They note that the potential wells for the regimes may be too shallow in the models. \textcolor{black}{Addionally \cite{falkena2022detection} found that using more regimes improved the representation of wintertime Euro-Atlantic sector dynamics.} We posit that this multimodality, in breaking the ergodic assumption necessary for the ensemble mean and variance, would consequently invalidate interpretation of those statistics, such as RPC. In fact, it leads one to reach the same conclusion as \cite{https://doi.org/10.1002/qj.4440}, the signal-to-noise ``paradox" is not really paradoxical; because ergodically we cannot expect the ensemble mean/variance to be the most likely phase-space trajectory.

Whilst our approach using the ergodic theorem to ensembles is new to this problem, we note that Lorenz discussed transitive and intransitive systems and climate, see \cite{Lorenz1968}, \cite{LORENZ1976495} and \cite{https://doi.org/10.1034/j.1600-0870.1990.t01-2-00005.x}. Ergodicity and transitivity are essentially the same mathematical phenomenon (see pedagogical discussion by \cite{maths_trans}). Lorenz's discussion of transitivity sets up in more general terms the concepts for climatic regime states which were later more clearly defined by \cite{ANonlinearDynamicalPerspectiveonClimatePrediction}. The ergodic theorem therefore provides further supporting rigour and a different perspective on the multi-modality paradigm.

One mitigation strategy for the multi-modality issue is to increase the ensemble size. However, even with a relatively large ensemble/\textcolor{black}{sample}, the paradox has been found to persist, e.g. see \cite{https://doi.org/10.1002/asl.1212} \textcolor{black}{and \cite{shi2015impact}}. There is weak evidence that the paradox exists on longer than multi-decadal timescales \cite{Scaife2018}. The reason for this may be understood from an ergodic perspective, as to define a mean climate state the interval $\Delta T$ over which we compute the average, is sufficiently long that the distributions $P^i$ converge \cite{Tantet_2016}. One could increase the averaging window to see how this effects the RPC. However one cannot increase it too much else the definition of the observable would be changed. In other words, for very large averaging windows, the observable simply becomes a different kind of climatic average: sub-seasonal to seasonal, seasonal to annual etc.

Another way in which failure of ergodicity can occur is if the ensemble members are not exchangeable with each other. This can be seen from equation \ref{ens_mean}, as one cannot factorise a single $P^i(x_t)$ in the integrand unless they are all \textit{perfectly} exchangeable. According to \cite{https://doi.org/10.1002/qj.3387}, the ensemble members of the operational ECMWF model, which has reported anomalous RPCs, are not truly exchangeable because its initial perturbations have a plus-minus symmetry \textcolor{black}{(i.e., initial perturbation of member $2k$ is minus the perturbation of member $(2k - 1)$), introducing systematic differences between ensemble members.} Other forecasting systems which are known to not satisfy exchangeability include the Meteo-France global ensemble, and Canadian ensemble prediction system, see \cite{https://doi.org/10.1002/qj.3094}. Also note that \cite{Siegert2016}, see section 4b, raised the possibility that the Met Office's GloSea5 climate prediction system ensemble members might not be exactly exchangeable. As reported by \cite{antje_report}, it is understood that exchangeability of observations and ensemble members provides a strong criterion for the signal-and-noise paradox. What is not recognised explicitly is that the members themselves must also be exchangeable to avoid failure of the ergodic theorem when interpreting the ensemble mean/variance as the most likely trajectory and then using these for the RPC. Combined with multi-modality, non-exchangeability could make it more likely for situations of $V(e)<V(o)$ to occur as we need to correctly weight each member according to its uniquely evolved PDF. Members might appear ``stuck" - in contrast to their evolved PDFs, which if used as weights would make the ensemble mean and variance better estimates of the most likely phase-space trajectory.

If we look at other simpler physical systems we readily find cases where failure of the ergodic assumption leads to significant incorrect statistical measures. Examples of such systems include supercooled liquids to glass transitions (see \cite{PhysRevA.39.3563}), diffusive processes within the plasma membrane of living cells (see \cite{Weigel2011-fn}) or general noise processes (see \cite{doi:10.1098/rsif.2022.0095}). In each of these cases, handling ensemble averages without due attention to the ergodic assumption leads to in simple terms, wrong answers. \cite{PhysRevLett.110.100603} demonstrates an example of this in geometric brownian motion where nonergodicity can lead to the ensemble mean growing exponentially, whilst simultaneously any individual trajectory decays exponentially according to its time average.

\section{Suggestions for future studies}
Future work to examine the hypotheses in this Letter could include testing the Thirumalai–Mountain effective ergodic convergence metric, $\Omega(t)$ measures the effective ergodicity by the difference between the time average of an observable and its ensemble average over the entire system. So for an ensemble of size $n$,
\begin{equation}
\Omega(t) = \frac{1}{n} \sum_{i=1}^n\left[f(t)_i-E(f(t))\right]^2,
\end{equation}
where the ensemble average is as defined by equation \ref{eqn:ensemble_mean}. In the long-time limit $\Omega \rightarrow 0$ for an ergodic system. The rate at which this occurs can be used to indicate timescales over which an ergodic approximation might be appropriate. This metric has, as far as we are aware, not been tested before in ensemble prediction systems. However, it is utilised for earthquake forecasting see e.g. \cite{PhysRevLett.91.238501} and \cite{PhysRevE.75.066107}, and studies of fluids, see \cite{10.1063/1.2035080}. For other metrics related to ergodicity see \cite{MATHEW2011432}.

Given sufficient computing resources, one could develop an ``ensemble-of-ensembles" approach to test perfect exchangeability. Once could generate at each time step \textcolor{black}{small} perturbations to each member generating an ensemble per member, so as to compute finite time Lyapunov exponents. From this one could compute the Lyapunov spectrum per main member. If the spectrum varies between the ensemble members then this would imply that fundamentally the members evolve into different state spaces, and thus are not exchangeable, hence failure of the ergodic mean to represent the most likely trajectory.

An approach to relax the ergodicity constraint but still make use of ensembles is to enforce dynamical invariants. These are quantities that are conserved along the phase-space path of the system. For example, in the probability theory for non-equilibrium gravitational systems developed by \cite{10.1093/mnras/stv1146}, dynamical invariants are employed with the explicit purpose of not relying ergodicity assumptions. For turbulent fluids, the most obvious dynamical invariant would be the Lyapunov spectrum and fractal dimension, besides vorticity. Recently \cite{10.1063/5.0156999} developed such a scheme for a numerical weather model machine learning training method. They tested their model on the classic Lorenz 1996 chaotic dynamical system, and found including ergodic constraints improved forecast skill.

\section{Summary}
The signal-to-noise paradox, assessed by the RPC metric, relies on interpreting ensemble mean and variance as the most probable phase-space trajectory. However, this interpretation, relies on the ergodic theorem, which faces potential failure due to insufficient ensemble size, short averaging windows, and non-exchangeability of members. In cases like winter NAO forecasts, evidence of multi-modal regime behaviour \textcolor{black}{or intransitivity between NAO phases, irrespective of multi-modality}, could exacerbate this failure, leading to hindered exploration of the complete phase space. Consequently, the ensemble mean and variance cannot reliably represent the most likely trajectory, resulting in the paradox when the ensemble becomes stuck - correlating with observations but underestimating the variance.

\section*{ACKNOWLEDGEMENTS}
D.J.B. is funded by the UK Science and Technology Facilities Council grant ST/W507441/1. He is particularly grateful to Professor Adam Scaife who introduced him to the problem, as well as Dr Jose Rodriguez and Dr Leon Hermanson of the Met Office and Professor Jorge Peñarrubia of The University of Edinburgh, for giving their time to insightful tête-à-tête discussions. He also thanks Alexander Tully of The University of Liverpool for comments on the manuscript. D.J.B. expresses his gratitude to two anonymous reviewers and Associate Editor Dr. Steven Hardiman for their invaluable comments and feedback. For the purpose of open access, the author has applied a Creative Commons Attribution (CC BY) licence to any Author Accepted Manuscript version arising from this submission.

\section*{CONFLICT OF INTEREST}
The author declares no conflict of interest.

\section*{DATA AVAILABILITY STATEMENT}
No data was used or produced in the course of this research.

\printendnotes
\bibliography{main}

\begin{thebibliography}{55}
\expandafter\ifx\csname natexlab\endcsname\relax\def\natexlab#1{#1}\fi
\expandafter\ifx\csname url\endcsname\relax
  \def\url#1{\texttt{#1}}\fi
\expandafter\ifx\csname urlprefix\endcsname\relax\def\urlprefix{URL: }\fi

\bibitem[{Ahmed et~al.(2022)Ahmed, Chumley, Cook, Cox, Grant, Petela, Rothrock and Xhafaj}]{ahmed2022dynamics}
Ahmed, J., Chumley, T., Cook, S., Cox, C., Grant, H., Petela, N., Rothrock, B. and Xhafaj, R. (2022) Dynamics of the no-slip galton board.
\newblock \urlprefix\url{https://arxiv.org/abs/2208.07790}.

\bibitem[{Birkhoff(1931)}]{1bd409ba-8905-3534-9c6f-e5a5e359481e}
Birkhoff, G.~D. (1931) Proof of the ergodic theorem.
\newblock \textit{Proceedings of the National Academy of Sciences of the United States of America}, \textbf{17}, 656--660.
\newblock \urlprefix\url{http://www.jstor.org/stable/86016}.

\bibitem[{Bröcker et~al.(2023)Bröcker, Charlton–Perez and Weisheimer}]{https://doi.org/10.1002/qj.4440}
Bröcker, J., Charlton–Perez, A.~J. and Weisheimer, A. (2023) A statistical perspective on the signal-to-noise paradox.
\newblock \textit{Quarterly Journal of the Royal Meteorological Society}, \textbf{149}, 911--923.
\newblock \urlprefix\url{https://rmets.onlinelibrary.wiley.com/doi/abs/10.1002/qj.4440}.

\bibitem[{Charlton-Perez et~al.(2019)Charlton-Perez, Br{\"o}cker, Stockdale and Johnson}]{Charlton-Perez2019}
Charlton-Perez, A., Br{\"o}cker, J., Stockdale, T.~N. and Johnson, S.~J. (2019) When and where do ecmwf seasonal forecast systems exhibit anomalously low signal-to-noise ratio?
\newblock \textit{Quarterly Journal of the Royal Meteorological Society}, \textbf{145}, 3466--3478.
\newblock \urlprefix\url{https://doi.org/10.1002/qj.3631}.

\bibitem[{Cottrell et~al.(2024)Cottrell, Screen and Scaife}]{https://doi.org/10.1002/asl.1212}
Cottrell, F.~M., Screen, J.~A. and Scaife, A.~A. (2024) Signal-to-noise errors in free-running atmospheric simulations and their dependence on model resolution.
\newblock \textit{Atmospheric Science Letters}, \textbf{25}, e1212.
\newblock \urlprefix\url{https://rmets.onlinelibrary.wiley.com/doi/abs/10.1002/asl.1212}.

\bibitem[{Dunstone et~al.(2023)Dunstone, Smith, Hardiman, Hermanson, Ineson, Kay, Li, Lockwood, Scaife, Thornton, Ting and Wang}]{Dunstone2023}
Dunstone, N., Smith, D.~M., Hardiman, S.~C., Hermanson, L., Ineson, S., Kay, G., Li, C., Lockwood, J.~F., Scaife, A.~A., Thornton, H., Ting, M. and Wang, L. (2023) Skilful predictions of the summer north atlantic oscillation.
\newblock \textit{Communications Earth {\&} Environment}, \textbf{4}, 409.
\newblock \urlprefix\url{https://doi.org/10.1038/s43247-023-01063-2}.

\bibitem[{Eade et~al.(2014)Eade, Smith, Scaife, Wallace, Dunstone, Hermanson and Robinson}]{Eade2014}
Eade, R., Smith, D., Scaife, A.~A., Wallace, E., Dunstone, N., Hermanson, L. and Robinson, N. (2014) Do seasonal-to-decadal climate predictions underestimate the predictability of the real world?
\newblock \textit{Geophysical research letters}, \textbf{41}, 5620--5628.
\newblock \urlprefix\url{https://doi.org/10.1002/2014gl061146}.

\bibitem[{Eckmann and Ruelle(1985)}]{RevModPhys.57.617}
Eckmann, J.~P. and Ruelle, D. (1985) Ergodic theory of chaos and strange attractors.
\newblock \textit{Rev. Mod. Phys.}, \textbf{57}, 617--656.
\newblock \urlprefix\url{https://link.aps.org/doi/10.1103/RevModPhys.57.617}.

\bibitem[{Epstein(1969)}]{epstein1969stochastic}
Epstein, E.~S. (1969) Stochastic dynamic prediction.
\newblock \textit{Tellus}, \textbf{21}, 739--759.

\bibitem[{Falkena et~al.(2022)Falkena, de~Wiljes, Weisheimer and Shepherd}]{falkena2022detection}
Falkena, S.~K., de~Wiljes, J., Weisheimer, A. and Shepherd, T.~G. (2022) {Detection of interannual ensemble forecast signals over the North Atlantic and Europe using atmospheric circulation regimes}.
\newblock \textit{Quarterly Journal of the Royal Meteorological Society}, \textbf{148}, 434--453.
\newblock \urlprefix\url{https://rmets.onlinelibrary.wiley.com/doi/abs/10.1002/qj.4213}.

\bibitem[{Galton(1889)}]{galton1889natural}
Galton, F. (1889) \textit{Natural Inheritance}.
\newblock No. v. 42; v. 590 in Natural Inheritance. Macmillan.

\bibitem[{Hardiman et~al.(2022)Hardiman, Dunstone, Scaife, Smith, Comer, Nie and Ren}]{hardiman2022missing}
Hardiman, S.~C., Dunstone, N.~J., Scaife, A.~A., Smith, D.~M., Comer, R., Nie, Y. and Ren, H.-L. (2022) Missing eddy feedback may explain weak signal-to-noise ratios in climate predictions.
\newblock \textit{npj Climate and Atmospheric Science}, \textbf{5}, 57.

\bibitem[{Hoover and Moran(1992)}]{Hoover1992-az}
Hoover, W.~G. and Moran, B. (1992) Viscous attractor for the galton board.
\newblock \textit{Chaos}, \textbf{2}, 599--602.

\bibitem[{Judd(2007)}]{judd2007galton}
Judd, K. (2007) Galton's quincunx: Random walk or chaos?
\newblock \textit{International Journal of Bifurcation and Chaos}, \textbf{17}, 4463--4469.

\bibitem[{Knight et~al.(2022)Knight, Scaife and Maidens}]{https://doi.org/10.1029/2022GL100471}
Knight, J.~R., Scaife, A.~A. and Maidens, A. (2022) An extratropical contribution to the signal-to-noise paradox in seasonal climate prediction.
\newblock \textit{Geophysical Research Letters}, \textbf{49}, e2022GL100471.
\newblock \urlprefix\url{https://agupubs.onlinelibrary.wiley.com/doi/abs/10.1029/2022GL100471}.
\newblock E2022GL100471 2022GL100471.

\bibitem[{Kumar(2009)}]{kumar2009finite}
Kumar, A. (2009) Finite samples and uncertainty estimates for skill measures for seasonal prediction.
\newblock \textit{Monthly Weather Review}, \textbf{137}, 2622--2631.

\bibitem[{{Kumi{\v{c}}{\'a}k}(2000)}]{2000AIPC..511..144K}
{Kumi{\v{c}}{\'a}k}, J. (2000) {Galton board as a model for fluctuations}.
\newblock In \textit{Unsolved Problems of Noise and Fluctuations: UPoN'99: Second International Conference}, vol. 511 of \textit{American Institute of Physics Conference Series}, 144--149.

\bibitem[{Leutbecher(2019)}]{https://doi.org/10.1002/qj.3387}
Leutbecher, M. (2019) Ensemble size: How suboptimal is less than infinity?
\newblock \textit{Quarterly Journal of the Royal Meteorological Society}, \textbf{145}, 107--128.
\newblock \urlprefix\url{https://rmets.onlinelibrary.wiley.com/doi/abs/10.1002/qj.3387}.

\bibitem[{Leutbecher et~al.(2017)Leutbecher, Lock, Ollinaho, Lang, Balsamo, Bechtold, Bonavita, Christensen, Diamantakis, Dutra, English, Fisher, Forbes, Goddard, Haiden, Hogan, Juricke, Lawrence, MacLeod, Magnusson, Malardel, Massart, Sandu, Smolarkiewicz, Subramanian, Vitart, Wedi and Weisheimer}]{https://doi.org/10.1002/qj.3094}
Leutbecher, M., Lock, S.-J., Ollinaho, P., Lang, S. T.~K., Balsamo, G., Bechtold, P., Bonavita, M., Christensen, H.~M., Diamantakis, M., Dutra, E., English, S., Fisher, M., Forbes, R.~M., Goddard, J., Haiden, T., Hogan, R.~J., Juricke, S., Lawrence, H., MacLeod, D., Magnusson, L., Malardel, S., Massart, S., Sandu, I., Smolarkiewicz, P.~K., Subramanian, A., Vitart, F., Wedi, N. and Weisheimer, A. (2017) Stochastic representations of model uncertainties at ecmwf: state of the art and future vision.
\newblock \textit{Quarterly Journal of the Royal Meteorological Society}, \textbf{143}, 2315--2339.
\newblock \urlprefix\url{https://rmets.onlinelibrary.wiley.com/doi/abs/10.1002/qj.3094}.

\bibitem[{Lorenz(1963)}]{lorenz1963deterministic}
Lorenz, E.~N. (1963) Deterministic nonperiodic flow.
\newblock \textit{Journal of atmospheric sciences}, \textbf{20}, 130--141.

\bibitem[{Lorenz(1968)}]{Lorenz1968}
--- (1968) \textit{Climatic Determinism}, 1--3.
\newblock Boston, MA: American Meteorological Society.
\newblock \urlprefix\url{https://doi.org/10.1007/978-1-935704-38-6_1}.

\bibitem[{Lorenz(1969)}]{lorenz1969predictability}
--- (1969) The predictability of a flow which possesses many scales of motion.
\newblock \textit{Tellus}, \textbf{21}, 289--307.

\bibitem[{Lorenz(1976)}]{LORENZ1976495}
--- (1976) Nondeterministic theories of climatic change.
\newblock \textit{Quaternary Research}, \textbf{6}, 495--506.
\newblock \urlprefix\url{https://www.sciencedirect.com/science/article/pii/0033589476900223}.

\bibitem[{Lorenz(1990)}]{https://doi.org/10.1034/j.1600-0870.1990.t01-2-00005.x}
--- (1990) Can chaos and intransitivity lead to interannual variability?
\newblock \textit{Tellus A}, \textbf{42}, 378--389.
\newblock \urlprefix\url{https://onlinelibrary.wiley.com/doi/abs/10.1034/j.1600-0870.1990.t01-2-00005.x}.

\bibitem[{Mangalam and Kelty-Stephen(2022)}]{doi:10.1098/rsif.2022.0095}
Mangalam, M. and Kelty-Stephen, D.~G. (2022) Ergodic descriptors of non-ergodic stochastic processes.
\newblock \textit{Journal of The Royal Society Interface}, \textbf{19}, 20220095.
\newblock \urlprefix\url{https://royalsocietypublishing.org/doi/abs/10.1098/rsif.2022.0095}.

\bibitem[{Mathew and Mezić(2011)}]{MATHEW2011432}
Mathew, G. and Mezić, I. (2011) Metrics for ergodicity and design of ergodic dynamics for multi-agent systems.
\newblock \textit{Physica D: Nonlinear Phenomena}, \textbf{240}, 432--442.
\newblock \urlprefix\url{https://www.sciencedirect.com/science/article/pii/S016727891000285X}.

\bibitem[{Murphy and Palmer(1986)}]{murphy1986experimental}
Murphy, J. and Palmer, T. (1986) Experimental monthly long-range forecasts for the united-kingdom. 2. a real-time long-range forecast by an ensemble of numerical integrations.
\newblock \textit{Meteorological Magazine}, \textbf{115}.

\bibitem[{Ollagnier(1985)}]{MoulinOllagnier1985}
Ollagnier, J.~M. (1985) \textit{Ergodic Theory and Statistical Mechanics}, vol. 1115 of \textit{Lecture Notes in Mathematics}.
\newblock Springer Berlin, Heidelberg.

\bibitem[{O'Reilly et~al.(2019)O'Reilly, Weisheimer, Woollings, Gray and MacLeod}]{https://doi.org/10.1002/qj.3413}
O'Reilly, C.~H., Weisheimer, A., Woollings, T., Gray, L.~J. and MacLeod, D. (2019) The importance of stratospheric initial conditions for winter north atlantic oscillation predictability and implications for the signal-to-noise paradox.
\newblock \textit{Quarterly Journal of the Royal Meteorological Society}, \textbf{145}, 131--146.
\newblock \urlprefix\url{https://rmets.onlinelibrary.wiley.com/doi/abs/10.1002/qj.3413}.

\bibitem[{Oss{\'o} et~al.(2020)Oss{\'o}, Sutton, Shaffrey and Dong}]{osso2020development}
Oss{\'o}, A., Sutton, R., Shaffrey, L. and Dong, B. (2020) Development, amplification, and decay of atlantic/european summer weather patterns linked to spring north atlantic sea surface temperatures.
\newblock \textit{Journal of Climate}, \textbf{33}, 5939--5951.

\bibitem[{Palmer and Hagedorn(2006)}]{palmer2006predictability}
Palmer, T. and Hagedorn, R. (2006) \textit{Predictability of Weather and Climate}.
\newblock Cambridge University Press.

\bibitem[{Palmer(1999)}]{ANonlinearDynamicalPerspectiveonClimatePrediction}
Palmer, T.~N. (1999) A nonlinear dynamical perspective on climate prediction.
\newblock \textit{Journal of Climate}, \textbf{12}, 575 -- 591.
\newblock \urlprefix\url{https://journals.ametsoc.org/view/journals/clim/12/2/1520-0442_1999_012_0575_andpoc_2.0.co_2.xml}.

\bibitem[{Peters and Klein(2013)}]{PhysRevLett.110.100603}
Peters, O. and Klein, W. (2013) Ergodicity breaking in geometric brownian motion.
\newblock \textit{Phys. Rev. Lett.}, \textbf{110}, 100603.
\newblock \urlprefix\url{https://link.aps.org/doi/10.1103/PhysRevLett.110.100603}.

\bibitem[{Peñarrubia(2015)}]{10.1093/mnras/stv1146}
Peñarrubia, J. (2015) {A probability theory for non-equilibrium gravitational systems}.
\newblock \textit{Monthly Notices of the Royal Astronomical Society}, \textbf{451}, 3537--3550.
\newblock \urlprefix\url{https://doi.org/10.1093/mnras/stv1146}.

\bibitem[{Platt et~al.(2023)Platt, Penny, Smith, Chen and Abarbanel}]{10.1063/5.0156999}
Platt, J.~A., Penny, S.~G., Smith, T.~A., Chen, T.-C. and Abarbanel, H. D.~I. (2023) {Constraining chaos: Enforcing dynamical invariants in the training of reservoir computers}.
\newblock \textit{Chaos: An Interdisciplinary Journal of Nonlinear Science}, \textbf{33}, 103107.
\newblock \urlprefix\url{https://doi.org/10.1063/5.0156999}.

\bibitem[{Scaife et~al.(2014)Scaife, Arribas, Blockley, Brookshaw, Clark, Dunstone, Eade, Fereday, Folland, Gordon, Hermanson, Knight, Lea, MacLachlan, Maidens, Martin, Peterson, Smith, Vellinga, Wallace, Waters and Williams}]{Scaife2014}
Scaife, A.~A., Arribas, A., Blockley, E.~W., Brookshaw, A., Clark, R.~T., Dunstone, N., Eade, R., Fereday, D., Folland, C.~K., Gordon, M., Hermanson, L., Knight, J., Lea, D.~J., MacLachlan, C., Maidens, A., Martin, M., Peterson, A., Smith, D., Vellinga, M., Wallace, E., Waters, J. and Williams, A.~J. (2014) Skillful long‐range prediction of european and north american winters.
\newblock \textit{Geophysical Research Letters}, \textbf{41}, 2514--2519.
\newblock \urlprefix\url{https://doi.org/10.1002/2014gl059637}.

\bibitem[{Scaife et~al.(2019)Scaife, Camp, Comer, Davis, Dunstone, Gordon, MacLachlan, Martin, Nie, Ren et~al.}]{scaife2019does}
Scaife, A.~A., Camp, J., Comer, R., Davis, P., Dunstone, N., Gordon, M., MacLachlan, C., Martin, N., Nie, Y., Ren, H.-L. et~al. (2019) Does increased atmospheric resolution improve seasonal climate predictions?
\newblock \textit{Atmospheric Science Letters}, \textbf{20}, e922.

\bibitem[{Scaife and Smith(2018)}]{Scaife2018}
Scaife, A.~A. and Smith, D. (2018) {A signal-to-noise paradox in climate science}.
\newblock \textit{npj Climate and Atmospheric Science}, \textbf{1}, 28.
\newblock \urlprefix\url{https://doi.org/10.1038/s41612-018-0038-4}.

\bibitem[{Shalizi(2007)}]{maths_trans}
Shalizi, C. (2007) Advanced probability ii or almost none of the theory of stochastic processes.
\newblock \urlprefix\url{https://www.stat.cmu.edu/~cshalizi/754/notes/all.pdf}.
\newblock [Online; accessed 5 July 2024].

\bibitem[{Shi et~al.(2015)Shi, Schaller, MacLeod, Palmer and Weisheimer}]{shi2015impact}
Shi, W., Schaller, N., MacLeod, D., Palmer, T.~N. and Weisheimer, A. (2015) {Impact of hindcast length on estimates of seasonal climate predictability}.
\newblock \textit{Geophysical Research Letters}, \textbf{42}, 1554--1559.
\newblock \urlprefix\url{https://agupubs.onlinelibrary.wiley.com/doi/abs/10.1002/2014GL062829}.

\bibitem[{Siegert et~al.(2016)Siegert, Stephenson, Sansom, Scaife, Eade and Arribas}]{Siegert2016}
Siegert, S., Stephenson, D.~B., Sansom, P.~G., Scaife, A.~A., Eade, R. and Arribas, A. (2016) A bayesian framework for verification and recalibration of ensemble forecasts: How uncertain is nao predictability?
\newblock \textit{Journal of Climate}, \textbf{29}, 995--1012.
\newblock \urlprefix\url{https://doi.org/10.1175/jcli-d-15-0196.1}.

\bibitem[{de~Souza and Wales(2005)}]{10.1063/1.2035080}
de~Souza, V.~K. and Wales, D.~J. (2005) {Diagnosing broken ergodicity using an energy fluctuation metric}.
\newblock \textit{The Journal of Chemical Physics}, \textbf{123}, 134504.
\newblock \urlprefix\url{https://doi.org/10.1063/1.2035080}.

\bibitem[{Stockdale et~al.(2015)Stockdale, Molteni and Ferranti}]{Stockdale2015}
Stockdale, T.~N., Molteni, F. and Ferranti, L. (2015) Atmospheric initial conditions and the predictability of the arctic oscillation.
\newblock \textit{Geophysical Research Letters}, \textbf{42}, 1173--1179.
\newblock \urlprefix\url{https://doi.org/10.1002/2014gl062681}.

\bibitem[{Strommen(2020)}]{strommen2020jet}
Strommen, K. (2020) Jet latitude regimes and the predictability of the north atlantic oscillation.
\newblock \textit{Quarterly Journal of the Royal Meteorological Society}, \textbf{146}, 2368--2391.

\bibitem[{Strommen et~al.(2023)Strommen, MacRae and Christensen}]{https://doi.org/10.1029/2023GL103710}
Strommen, K., MacRae, M. and Christensen, H. (2023) On the relationship between reliability diagrams and the “signal-to-noise paradox”.
\newblock \textit{Geophysical Research Letters}, \textbf{50}, e2023GL103710.
\newblock \urlprefix\url{https://agupubs.onlinelibrary.wiley.com/doi/abs/10.1029/2023GL103710}.
\newblock E2023GL103710 2023GL103710.

\bibitem[{Strommen and Palmer(2018)}]{Strommen2018}
Strommen, K. and Palmer, T. (2018) Signal and noise in regime systems: A hypothesis on the predictability of the north atlantic oscillation.
\newblock \textit{Quarterly Journal of the Royal Meteorological Society}, \textbf{145}, 147--163.
\newblock \urlprefix\url{https://doi.org/10.1002/qj.3414}.

\bibitem[{Tantet(2016)}]{Tantet_2016}
Tantet, A. J.~J. (2016) \textit{Ergodic theory of climate: variability, stability and response}.
\newblock Ph.D. thesis, Utrecht University.

\bibitem[{Thirumalai et~al.(1989)Thirumalai, Mountain and Kirkpatrick}]{PhysRevA.39.3563}
Thirumalai, D., Mountain, R.~D. and Kirkpatrick, T.~R. (1989) Ergodic behavior in supercooled liquids and in glasses.
\newblock \textit{Phys. Rev. A}, \textbf{39}, 3563--3574.
\newblock \urlprefix\url{https://link.aps.org/doi/10.1103/PhysRevA.39.3563}.

\bibitem[{Tiampo et~al.(2007)Tiampo, Rundle, Klein, Holliday, S\'a~Martins and Ferguson}]{PhysRevE.75.066107}
Tiampo, K.~F., Rundle, J.~B., Klein, W., Holliday, J., S\'a~Martins, J.~S. and Ferguson, C.~D. (2007) Ergodicity in natural earthquake fault networks.
\newblock \textit{Phys. Rev. E}, \textbf{75}, 066107.
\newblock \urlprefix\url{https://link.aps.org/doi/10.1103/PhysRevE.75.066107}.

\bibitem[{Tiampo et~al.(2003)Tiampo, Rundle, Klein, Martins and Ferguson}]{PhysRevLett.91.238501}
Tiampo, K.~F., Rundle, J.~B., Klein, W., Martins, J. S.~S. and Ferguson, C.~D. (2003) Ergodic dynamics in a natural threshold system.
\newblock \textit{Phys. Rev. Lett.}, \textbf{91}, 238501.
\newblock \urlprefix\url{https://link.aps.org/doi/10.1103/PhysRevLett.91.238501}.

\bibitem[{Tsonis and Elsner(1989)}]{ChaosStrangeAttractorsandWeather}
Tsonis, A.~A. and Elsner, J.~B. (1989) Chaos, strange attractors, and weather.
\newblock \textit{Bulletin of the American Meteorological Society}, \textbf{70}, 14 -- 23.
\newblock \urlprefix\url{https://journals.ametsoc.org/view/journals/bams/70/1/1520-0477_1989_070_0014_csaaw_2_0_co_2.xml}.

\bibitem[{Weigel et~al.(2011)Weigel, Simon, Tamkun and Krapf}]{Weigel2011-fn}
Weigel, A.~V., Simon, B., Tamkun, M.~M. and Krapf, D. (2011) Ergodic and nonergodic processes coexist in the plasma membrane as observed by single-molecule tracking.
\newblock \textit{Proc Natl Acad Sci U S A}, \textbf{108}, 6438--6443.

\bibitem[{Weisheimer et~al.(2024)Weisheimer, Baker, Bröcker, Garfinkel, Hardiman, Hodson, Palmer, Robson, Scaife, Screen, Shepherd, Smith and Sutton}]{antje_report}
Weisheimer, A., Baker, L.~H., Bröcker, J., Garfinkel, C.~I., Hardiman, S.~C., Hodson, D. L.~R., Palmer, T.~N., Robson, J.~I., Scaife, A.~A., Screen, J.~A., Shepherd, T.~G., Smith, D.~M. and Sutton, R.~T. (2024) The signal-to-noise paradox in climate forecasts: Revisiting our understanding and identifying future priorities.
\newblock \textit{Bulletin of the American Meteorological Society}, \textbf{105}, E651 -- E659.
\newblock \urlprefix\url{https://journals.ametsoc.org/view/journals/bams/105/3/BAMS-D-24-0019.1.xml}.

\bibitem[{Weisheimer et~al.(2019)Weisheimer, Decremer, MacLeod, O'Reilly, Stockdale, Johnson and Palmer}]{Weisheimer2019}
Weisheimer, A., Decremer, D., MacLeod, D., O'Reilly, C.~H., Stockdale, T.~N., Johnson, S.~J. and Palmer, T. (2019) How confident are predictability estimates of the winter north atlantic oscillation.
\newblock \textit{Quarterly Journal of the Royal Meteorological Society}, \textbf{145}, 140--159.
\newblock \urlprefix\url{https://doi.org/10.1002/qj.3446}.

\bibitem[{Zhang(2019)}]{Zhang2019}
Zhang, Wei;~Kirtman, B.~P. (2019) Understanding the signal‐to‐noise paradox with a simple markov model.
\newblock \textit{Geophysical Research Letters}, \textbf{46}, 13308--13317.
\newblock \urlprefix\url{https://doi.org/10.1029/2019gl085159}.

\end{thebibliography}

\end{document}